\documentclass[twocolumn,tighten]{aastex63}
\hypersetup{linkcolor=red,citecolor=blue,filecolor=cyan,urlcolor=magenta}
\usepackage{amsmath}
\usepackage{natbib}

\newcommand{\hdone}{\object[HD 196944]{HD~196944}}
\newcommand{\hdtwo}{\object[HD 222925]{HD~222925}}
\newcommand{\hdnine}{\object[HD 94028]{HD~94028}}

\newcommand{\loggf}{\mbox{$\log(gf)$}}
\newcommand{\kmsec}{\mbox{km~s$^{\rm -1}$}}
\newcommand{\logg}{\mbox{log~{\it g}}}

\newcommand{\teff}{\mbox{$T_{\rm eff}$}}
\newcommand{\vt}{\mbox{$v_{\rm t}$}}
\newcommand{\rpro}{\mbox{{\it r}-process}}
\newcommand{\spro}{\mbox{{\it s}-process}}
\newcommand{\ipro}{\mbox{{\it i}-process}}
\newcommand{\ncap}{\mbox{{\it n}-capture}}

\newcommand{\rtwo}{\mbox{\it r}-II}
\newcommand{\logeps}[1]{$\log\varepsilon$(#1)}
\newcommand{\cemps}{\mbox{CEMP-{\it s}}}

\shorttitle{Detection of Pb~\textsc{ii} in Metal-Poor Stars}
\shortauthors{Roederer et al.}
\accepted{for publication in the Astrophysical Journal Letters}

\begin{document}

\title{%
Detection of Pb~II in the Ultraviolet Spectra of
Three Metal-Poor Stars\footnote{%
Based on observations made with the NASA/ESA 
\textit{Hubble Space Telescope}, 
obtained at the Space Telescope Science Institute (STScI), which is 
operated by the Association of Universities for 
Research in Astronomy, Inc.\ (AURA) under NASA contract NAS~5-26555.
These observations are associated with programs 
GO-14161, GO-14765, and GO-15657.
This paper includes data taken at The McDonald Observatory
of The University of Texas at Austin.
}
}

\author{Ian U.\ Roederer}
\affiliation{%
Department of Astronomy, University of Michigan,
1085 S.\ University Ave., Ann Arbor, MI 48109, USA}
\affiliation{%
Joint Institute for Nuclear Astrophysics -- Center for the
Evolution of the Elements (JINA-CEE), USA}
\email{Email:\ iur@umich.edu}

\author{James E.\ Lawler}
\affiliation{%
Department of Physics, University of Wisconsin-Madison,
Madison, WI 53706, USA}

\author{Erika M.\ Holmbeck}
\affiliation{%
Center for Computational Relativity and Gravitation, 
Rochester Institute of Technology, Rochester, NY 14623, USA}
\affiliation{%
Joint Institute for Nuclear Astrophysics -- Center for the
Evolution of the Elements (JINA-CEE), USA}

\author{Timothy C.\ Beers}
\affiliation{%
Department of Physics, University of Notre Dame, 
Notre Dame, IN 46556, USA}
\affiliation{%
Joint Institute for Nuclear Astrophysics -- Center for the
Evolution of the Elements (JINA-CEE), USA}

\author{Rana Ezzeddine}
\affiliation{%
Department of Astronomy, University of Florida, Bryant Space Science Center,  
Gainesville, FL 32611, USA}
\affiliation{%
Joint Institute for Nuclear Astrophysics -- Center for the
Evolution of the Elements (JINA-CEE), USA}

\author{Anna Frebel}
\affiliation{%
Department of Physics and Kavli Institute for Astrophysics and Space Research, 
Massachusetts Institute of Technology, 
Cambridge, MA 02139, USA}
\affiliation{%
Joint Institute for Nuclear Astrophysics -- Center for the
Evolution of the Elements (JINA-CEE), USA}

\author{Terese T.\ Hansen}
\affiliation{%
George P.\ and Cynthia Woods Mitchell  
Institute for Fundamental Physics and Astronomy, 
Texas A\&M University, College Station, TX 77843, USA}
\affiliation{%
Department of Physics and Astronomy, Texas A\&M University,
College Station, TX 77843, USA}

\author{Inese I.\ Ivans}
\affiliation{%
Department of Physics and Astronomy, University of Utah,
Salt Lake City, UT 84112, USA}

\author{Amanda I.\ Karakas}
\affiliation{%
School of Physics and Astronomy, 
Monash University, VIC 3800, Australia}
\affiliation{%
ARC Centre of Excellence for All Sky Astrophysics in 3 Dimensions (ASTRO 3D)}

\author{Vinicius M.\ Placco}
\affiliation{%
NSF's Optical-Infrared Astronomy Research Laboratory, Tucson, AZ 85719, USA}
\affiliation{%
Joint Institute for Nuclear Astrophysics -- Center for the
Evolution of the Elements (JINA-CEE), USA}

\author{Charli M.\ Sakari}
\affiliation{%
Department of Physics and Astronomy, San Francisco State University,
San Francisco, CA 94132, USA}

\begin{abstract}

We report the first detection of the Pb~\textsc{ii} line at 2203.534~\AA\
in three metal-poor stars, using
ultraviolet spectra obtained with the
Space Telescope Imaging Spectrograph on board the 
\textit{Hubble Space Telescope}.
We perform a standard abundance analysis assuming
local thermodynamic equilibrium (LTE) 
to derive lead (Pb, $Z =$~82) abundances.
The Pb~\textsc{ii} line yields a higher abundance than
Pb~\textsc{i} lines by 
$+$0.36 $\pm$~0.34~dex and $+$0.49 $\pm$~0.28~dex 
in the stars 
\mbox{HD 94028} 
and 
\mbox{HD 196944},
where Pb~\textsc{i} lines had been detected previously.
The Pb~\textsc{ii} line is likely formed in LTE, and
these offsets affirm previous calculations showing
that Pb~\textsc{i} lines commonly used as abundance indicators
underestimate the Pb abundance in LTE.~
Pb is enhanced in the \spro-enriched stars
\mbox{HD 94028}
([Pb/Fe] $= +$0.95 $\pm$~0.14)
and
\mbox{HD 196944}
([Pb/Fe] $= +$2.28 $\pm$~0.23), and 
we show that $^{208}$Pb is the dominant Pb isotope in these two stars.
The \logeps{Pb/Eu} ratio in the \rpro-enhanced star
\mbox{HD 222925}
is 0.76~$\pm$~0.14, which matches the
Solar System \rpro\ ratio and
indicates that the Solar System \rpro\ residuals for
Pb are, in aggregate, correct.
The Th/Pb chronometer in 
\mbox{HD 222925} yields an age of 
8.2 $\pm$~5.8~Gyr, and 
we highlight the potential of the Th/Pb chronometer as a 
relatively model-insensitive age indicator in \rpro-enhanced stars.

\end{abstract}


\keywords{%
Nucleosynthesis (1131);
R-process (1324);
S-process (1419);
Stellar abundances (1577);
Ultraviolet astronomy (1736)
}

\section{Introduction}
\label{intro}

The element lead (Pb, $Z =$~82) has
fascinating nucleosynthetic origins.
Three of the four stable Pb isotopes 
($^{206}$Pb, $^{207}$Pb, and $^{208}$Pb) and
the one stable bismuth (Bi, $Z =$~83) 
isotope ($^{209}$Bi)
serve as both the high-mass termination point 
of the slow \ncap\ process (\spro) and 
the low-mass termination point of 
actinide $\alpha$-decay chains of radioactive isotopes
produced in the rapid \ncap\ process (\rpro).

Pb and Bi accumulate during the \spro\ as
\ncap\ and $\alpha$-decay
reactions cycle indefinitely
(e.g., \citealt{burbidge57}).
\citet{clayton67} recognized that most Pb in the Solar System
could not form through 
the \rpro\ or a ``smooth extension of the [\spro]
circumstances attending the synthesis for $A <$~200.''
\citet{gallino98} and \citet{travaglio01pb}
showed that the
high neutron-to-seed ratio that occurs in
low-metallicity stars during the 
asymptotic giant branch (AGB) phase of evolution
is responsible.
Most of that Pb is expected to be
$^{208}$Pb, which sits 
at both the $Z =$~82 proton shell closure and
the $N =$~126 neutron shell closure,
and its \ncap\ cross section is 
nearly an order of magnitude smaller
than that of neighboring nuclei
(cf.\ \citealt{ratzel04}).
Observations 
confirm enhanced Pb and Bi abundances,
relative to lighter \spro\ elements, 
in many \spro-enriched metal-poor stars
(e.g., \citealt{vaneck01,aoki02pb,ivans05}).

Pb is also a remarkable element in \rpro\ nucleosynthesis
because it is the final decay product for
most isotopes heavier than $^{209}$Bi,
including the three long-lived isotopes of the actinides
thorium (Th, $Z =$~90) and uranium (U, $Z =$~92):\
$^{232}$Th, 
$^{235}$U, and 
$^{238}$U,
which decay to 
$^{208}$Pb,
$^{207}$Pb, and
$^{206}$Pb, respectively.
The actinide elements can only be produced by \rpro\ nucleosynthesis,
yet our understanding of actinide production remains incomplete.
Pb abundances in metal-poor stars 
can bridge this gap in our understanding.
More than 85\% of Pb in old, metal-poor stars 
enhanced in \rpro\ elements
is formed through the decay of radioactive nuclei
with $A >$~209 \citep{cowan99},
so Pb abundances provide an important constraint on the
production of these isotopes
(e.g., \citealt{schatz02,wanajo02,eichler19}).

A few studies 
(e.g., \citealt{plez04})
have attempted to constrain model predictions by
assessing the \rpro\ contribution to Pb abundances,
but that work has been 
limited by observational uncertainties.
Only a few lines of Pb~\textsc{i} are detectable in the optical spectrum.
The most commonly-used line, at 4057.807~\AA,
is often weak and blended.
Furthermore, Pb is mostly ($\gtrsim$~99\%)
ionized in the atmospheres of metal-poor stars,
and departures from local thermodynamic equilibrium (LTE)
impact the Pb abundances derived from Pb~\textsc{i} lines
\citep{mashonkina12}.
These observational challenges can be overcome
by detecting Pb in its dominant ionization state, 
singly-ionized Pb.

In this Letter, we examine new and archival
ultraviolet (UV) spectra of three metal-poor stars
taken with the Space Telescope Imaging Spectrograph (STIS)
on board the \textit{Hubble Space Telescope}. 
These spectra show the
Pb~\textsc{ii} line at 2203.534~\AA, 
which is the only Pb~\textsc{ii} transition accessible
in near-UV, optical, or near-infrared
spectra of late-type stars.
This line has been observed previously in the spectra of
a few chemically-peculiar A-type stars
(e.g., \citealt{faraggiana89,cowley16}).
Previous attempts \citep{roederer14d} to detect the 
Pb~\textsc{ii} line 
in STIS E230M spectra ($R \equiv \lambda/\Delta\lambda$ = 30,000) 
of metal-poor stars have been unsuccessful.
Here, we present the first detection of this Pb~\textsc{ii} line
in metal-poor stars.

\section{Observations and Stellar Sample}
\label{sample}

STIS spectra of 
only three metal-poor stars yield compelling detections of the
Pb~\textsc{ii} line at 2203.534~\AA.~
These stars were selected for observations over the years
because they are bright, metal-poor, 
and in two cases
show extreme enhancements of \ncap\ elements.
UV spectra obtained with
STIS \citep{woodgate98}  
cover the Pb~\textsc{ii} line at 2203~\AA\
with the high resolving power 
($R$ = 114,000) of the
E230H grating, as summarized in Table~\ref{abundtab}.
Archival observations of the star \hdnine\
were downloaded from the 
Mikulski Archive for Space Telescopes (MAST).~
Two new sets of observations of the stars \hdone\
and \hdtwo\
were also downloaded through the MAST and 
processed automatically by the CALSTIS software package.
All spectra were shifted to a common velocity, co-added, and 
continuum normalized using IRAF.~
The signal-to-noise (S/N) ratios per pixel in the co-added spectra
are listed in Table~\ref{abundtab}.
These modest S/N ratios are sufficient to detect the lines
of interest because of the high resolving power.

\begin{deluxetable}{cccc}
\tablecaption{Log of Observations, Model Atmosphere Parameters, 
Metallicities, Ba, Eu, and Pb Abundances
\label{abundtab}}
\tabletypesize{\scriptsize}
\tablehead{
\colhead{Quantity} &
\colhead{HD 94028} &
\colhead{HD 196944} &
\colhead{HD 222925} 
}
\startdata
Prog.\ ID        & GO-14161                 & GO-14765                 & GO-15657                 \\
PI               & Peterson                 & Roederer                 & Roederer                 \\
Data Sets        & OCTKB0010-6030           & OD5A01010-14010          & ODX901010-60030          \\
$V_{\rm mag}$    & 8.22                     & 8.40                     & 9.03                     \\
S/N @ 2200~\AA   & 50/1                     & 40/1                     & 30/1                     \\
{\teff} (K)      &    6087\,$\pm$\,84   (1) &    5170\,$\pm$\,100  (2) &    5636\,$\pm$\,103  (3) \\
{\logg} [cgs]    &    4.37\,$\pm$\,0.13 (1) &    1.60\,$\pm$\,0.25 (2) &    2.54\,$\pm$\,0.17 (3) \\
{\vt} (\kmsec)   &    1.10\,$\pm$\,0.20 (1) &    1.55\,$\pm$\,0.10 (2) &    2.20\,$\pm$\,0.20 (3) \\
$V_{t}$ (\kmsec) &     1.6\,$\pm$\,0.3  (4) &     6.8\,$\pm$\,0.5  (5) &     7.0\,$\pm$\,0.5  (6) \\
{[M/H]}          & $-$1.60\,$\pm$\,0.10 (1) & $-$2.41\,$\pm$\,0.25 (2) & $-$1.50\,$\pm$\,0.10 (3) \\
{[Fe/H]}         & $-$1.65\,$\pm$\,0.08 (1) & $-$2.41\,$\pm$\,0.18 (2) & $-$1.46\,$\pm$\,0.08 (6) \\
{\logeps{Ba}}    &    1.06\,$\pm$\,0.11 (4) &    1.00\,$\pm$\,0.11 (2) &    1.26\,$\pm$\,0.09 (3) \\
{\logeps{Eu}}    & $-$0.62\,$\pm$\,0.13 (4) & $-$2.00\,$\pm$\,0.10 (2) &    0.38\,$\pm$\,0.09 (3) \\
{\logeps{Pb}}    &    1.34\,$\pm$\,0.12 (4) &    1.91\,$\pm$\,0.30 (4) &    1.14\,$\pm$\,0.16 (4) \\
{[Ba/Fe]}        & $+$0.53\,$\pm$\,0.09 (4) & $+$1.23\,$\pm$\,0.26 (2) & $+$0.54\,$\pm$\,0.06 (6) \\
{[Eu/Fe]}        & $+$0.51\,$\pm$\,0.13 (4) & $-$0.11\,$\pm$\,0.10 (2) & $+$1.32\,$\pm$\,0.08 (6) \\
{[Pb/Fe]}        & $+$0.95\,$\pm$\,0.14 (4) & $+$2.28\,$\pm$\,0.23 (4) & $+$0.56\,$\pm$\,0.14 (4) \\
\enddata
\tablecomments{%
We have rederived the barium (Ba, $Z =$~56; 
3 optical Ba~\textsc{ii} lines), 
europium (Eu, $Z =$~63; 2 optical Eu~\textsc{ii} lines), and 
Pb (1 UV Pb~\textsc{i} line)
abundances in \hdnine\
using our adopted model atmosphere
and a high-resolution optical spectrum
obtained using the Tull Coud\'{e} spectrograph
on the 2.7~m Harlan J.\ Smith Telescope at
McDonald Observatory
(see \citealt{roederer14c} for details).
These model parameters and 
the Ba and Eu $\log\varepsilon$ abundances 
are in agreement with those derived by \citet{peterson20}.
References are 
indicated by the numbers in parentheses:\
(1) \citet{roederer18b};
(2) \citet{placco15cemps};
(3) \citet{roederer18c};
(4) This study;
(5) \citet{roederer08a};
(6) Roederer et al.\ (in preparation).}
\end{deluxetable}

\hdnine\ 
shows moderate levels of 
enhancement of both \rpro\ and \spro\ elements
(\citealt{roederer16c,peterson20}; see Table~\ref{abundtab}).
\citeauthor{roederer16c}\ also found evidence that an
intermediate \ncap\ process (\ipro) may contribute to some of the
$Z <$~56 elements in \hdnine.
That study derived the Pb abundance from one Pb~\textsc{i} line
at 2833.054~\AA\ and 
concluded that the \spro\ dominates
the origin of the Pb in \hdnine.
There is no evidence from radial velocity (RV) measurements 
that \hdnine\ is in a binary system.

\hdone\ 
is one of the original ``Pb stars'' identified by
\citet{vaneck01}.
Numerous studies over the years have 
confirmed that \hdone\ is
a carbon-enhanced metal-poor
star enhanced in elements produced by the \spro\
(\cemps\ star; e.g., \citealt{zacs98,placco15cemps}).
The Pb abundance in \hdone\ has been derived previously
from Pb~\textsc{i} lines at 2833 and 4057~\AA.~
\hdone\ exhibits RV variations, and \citeauthor{placco15cemps}\
calculated an orbital period of 1325 $\pm$~12~d.

\hdtwo\ 
is a member of the class of highly \rpro-enhanced,
or \rtwo, stars (as defined in \citealt{holmbeck20}).
All heavy elements in \hdtwo\ were produced 
via \rpro\ nucleosynthesis \citep{roederer18c}.
Pb had not been detected previously in \hdtwo,
but \citeauthor{roederer18c}\ placed a tight upper limit on the
Pb abundance from the Pb~\textsc{i} line at 4057~\AA.~
\hdtwo\ shows no evidence of RV variations 
that would signal a stellar companion.

\section{Analysis}
\label{analysis}

\subsection{Atomic Data}
\label{pbatomic}

There are four stable isotopes of Pb:\
$^{204}$Pb (1.4\% in Solar System material),
$^{206}$Pb (24.1\%),
$^{207}$Pb (22.1\%), and
$^{208}$Pb (52.4\%).
Of these isotopes, $^{207}$Pb
has nonzero nuclear spin, $I = 1/2$, 
and thus has hyperfine splitting (HFS) structure.
The field shift, which results from the volume difference between
nuclei with the same number of protons but different numbers of neutrons,
also creates isotope shifts (IS).~
We adopt the ground level and excited level HFS $A$ values 
and the IS measurements 
from \citet{bouazza86}
for the Pb~\textsc{ii} line at 2203~\AA.~
We adopt the atomic transition probability
from \citet{quinet07},
\loggf\ = $-$0.24 $\pm$~0.07,
because it is normalized to a radiative lifetime measurement 
from laser induced fluorescence.
For comparison, the 
National Institutes of Standards and Technology (NIST)
Atomic Spectral Database (ASD; \citealt{kramida19})
recommends \loggf\ = $-$0.14,
based on theoretical calculations referenced there,
which is in fair agreement with our adopted value.
We present the line component pattern in Table~\ref{pbhfstab}.

\begin{deluxetable*}{cccccccc}
\tablecaption{Hyperfine Structure and Isotope Shifts
for the Pb~\textsc{ii} Line at 2203~\AA\
\label{pbhfstab}}
\tabletypesize{\small}
\tablehead{
\colhead{Wavenumber} &
\colhead{$\lambda_{\rm air}$} &
\colhead{$F_{\rm upper}$} &
\colhead{$F_{\rm lower}$} &
\colhead{Component Position} &
\colhead{Component Position} &
\colhead{Strength} &
\colhead{Isotope} \\
\colhead{(cm$^{-1}$)} &
\colhead{(\AA)} &
\colhead{} &
\colhead{} &
\colhead{(cm$^{-1}$)} &
\colhead{(\AA)} &
\colhead{} &
\colhead{}
}
\startdata
45367.486 & 2203.5342 & 0.5 & 1.5 & $-$0.175014 &    0.008501 & 1.000 & 204 \\
45367.486 & 2203.5342 & 0.5 & 1.5 & $-$0.067014 &    0.003255 & 1.000 & 206 \\
45367.486 & 2203.5342 & 1.0 & 2.0 &    0.061236 & $-$0.002975 & 0.625 & 207 \\
45367.486 & 2203.5342 & 1.0 & 1.0 &    0.095236 & $-$0.004626 & 0.125 & 207 \\
45367.486 & 2203.5342 & 0.0 & 1.0 & $-$0.252764 &    0.012278 & 0.250 & 207 \\
45367.486 & 2203.5342 & 0.5 & 1.5 &    0.040986 & $-$0.001991 & 1.000 & 208 \\
\enddata
\tablecomments{%
Energy levels from the NIST ASD and the index of air 
are used to compute the
center-of-gravity wavenumbers and air wavelengths, $\lambda_{\rm air}$, 
for a Solar System isotopic composition \citep{meija16}.
Line component positions are given relative to those values.
The strengths of each component are easily adjustable using 
Table~\ref{pbhfstab}
because a Solar System abundance pattern has not been assumed,
and strengths are normalized to sum to 1 for each isotope.
For example, the \loggf\ value of the $^{207}$Pb component with
$F_{\rm upper}$ = $F_{\rm lower}$ = 1.0 in a Solar System mix with 
$f_{207}$ = 0.221 would be 
$\log_{10}(0.221 \times 0.125 \times 10^{-0.24})$ = $-$1.80.
Table~\ref{pbhfstab} is available in the online edition
of the journal in machine-readable format.
}
\end{deluxetable*}

\subsection{Model Atmospheres}
\label{model}

We adopt the model parameters
(effective temperature, \teff;
log of the surface gravity, \logg;
microturbulent velocity, \vt;
model metallicity, [M/H])
derived previously for these three stars,
for consistency.
\hdnine\ is a main sequence dwarf, while
\hdone\ and \hdtwo\ are red horizontal branch stars.
We interpolate model atmospheres from the 1D,
$\alpha$-enhanced
ATLAS9 grid of models \citep{castelli04}.
Our synthetic spectra also include a 
macroturbulent velocity ($V_{t}$) component, 
which improves the fits to the high-resolution E230H spectra.
We derive $V_{t}$
by fitting the observed profiles of isolated lines of Fe-group elements.
These values are listed in Table~\ref{abundtab}.

\subsection{Pb Abundances}
\label{pbabunds}

\begin{figure}
\begin{center}
\includegraphics[angle=0,width=3.35in]{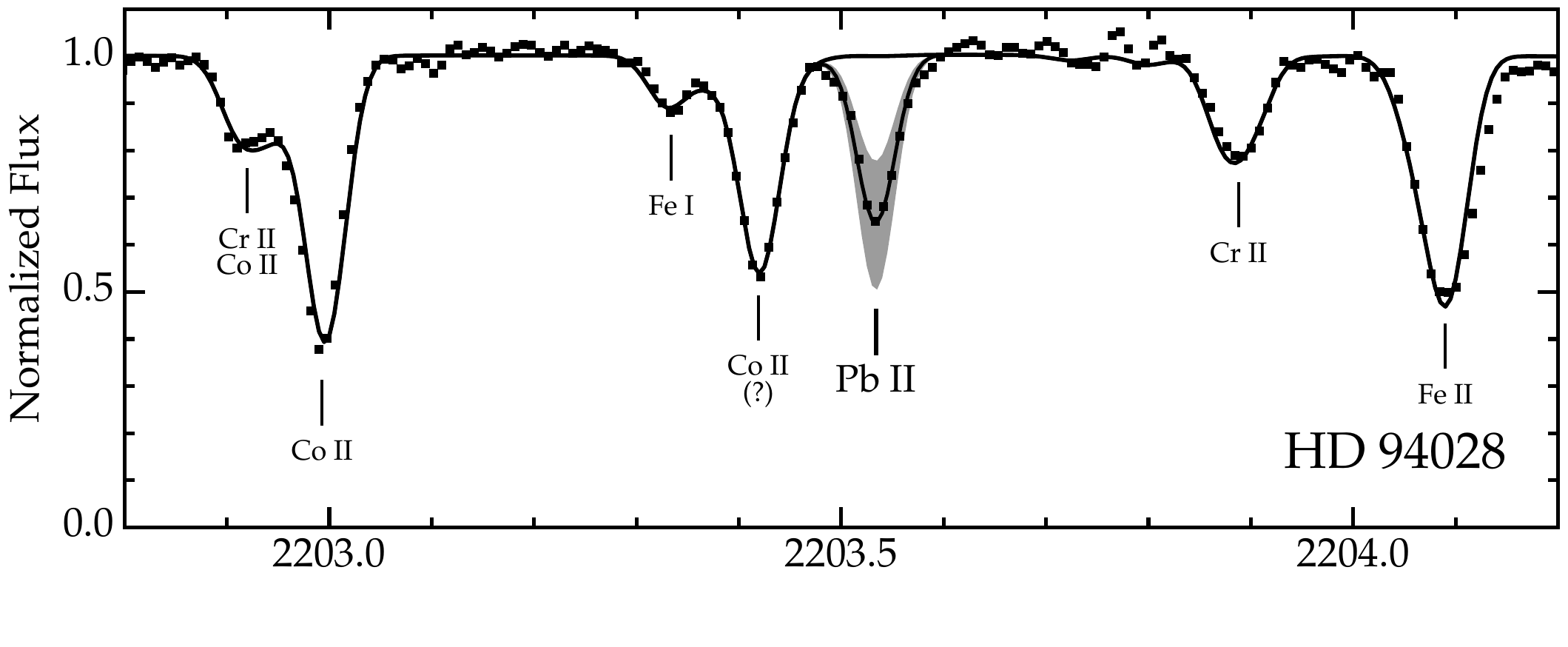} \\
\vspace*{-0.1in}
\includegraphics[angle=0,width=3.35in]{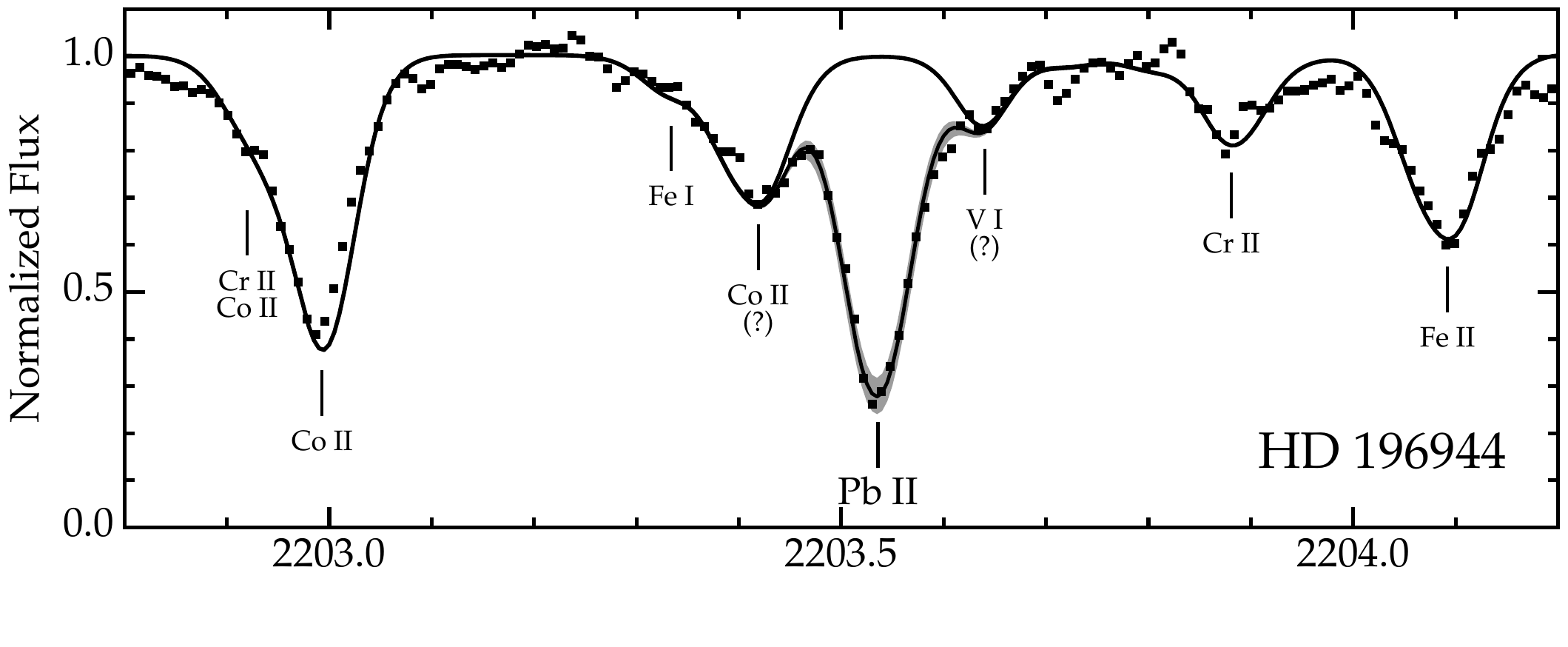} \\
\vspace*{-0.1in}
\includegraphics[angle=0,width=3.35in]{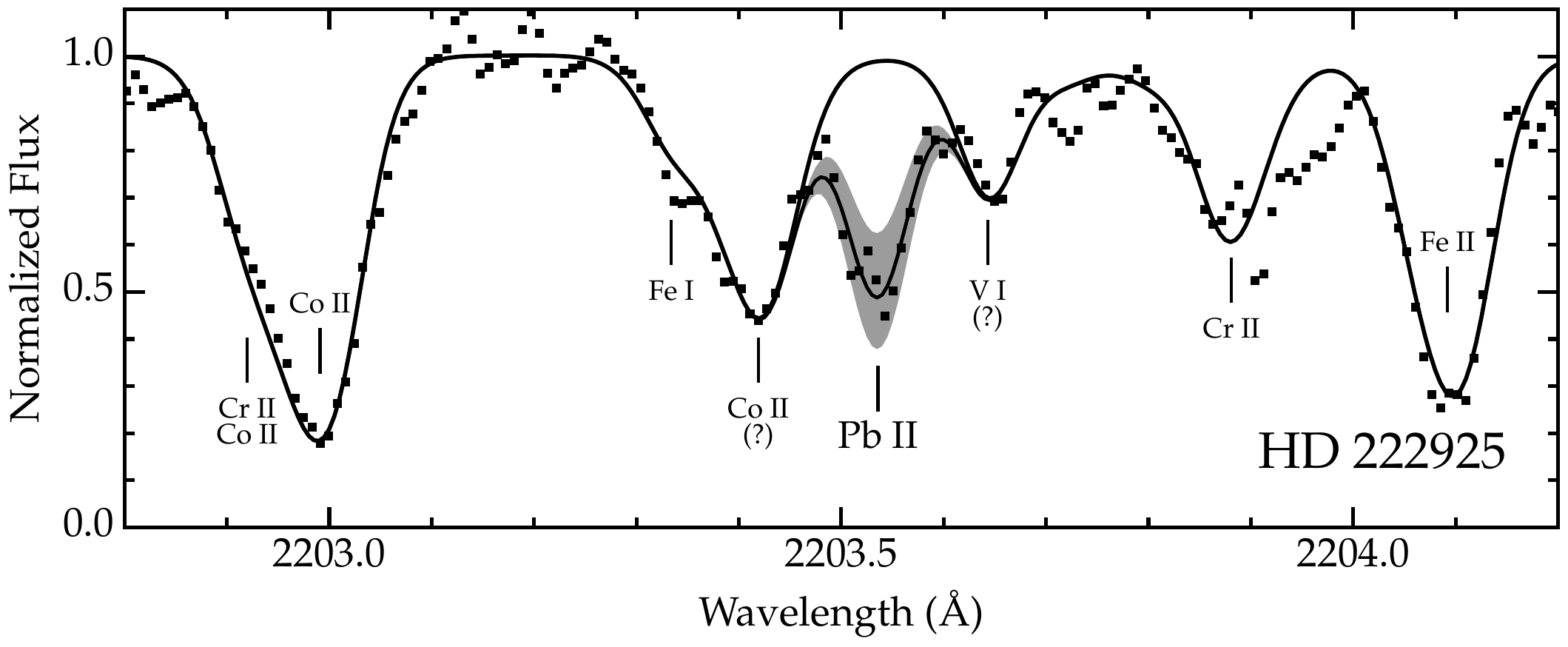}
\end{center}
\caption{
\label{specplot1}
Sections of the STIS/E230H spectra of \hdnine, \hdone, and \hdtwo\
around the Pb~\textsc{ii} line at 2203~\AA.~
The filled dots represent the observed spectrum.
The solid lines represent a synthetic spectrum with the 
best-fit abundance in each star,
and the gray bands represent a change in this abundance
by a factor of $\pm$~2 (0.3~dex).
The solid black line with no gray bands represents a synthetic spectrum 
with no contributions from Pb.
Other line identifications are marked.
}
\end{figure}

Figure~\ref{specplot1} illustrates the Pb~\textsc{ii} line
in the spectrum of each of the three stars.
Continuum regions around this line are easily identified,
and they are matched by the synthetic spectra.
We are confident that the absorption at 2203.534~\AA\
is due to Pb~\textsc{ii} for several reasons.
The line strength
varies with the expected heavy-element abundances in these
stars, not the abundances of iron-group elements that
are responsible for most UV absorption lines.
This line is---by many orders of magnitude---the strongest
Pb~\textsc{ii} line with $\lambda >$~2000~\AA,
so there is no expectation that other Pb~\textsc{ii} lines
could be detectable.
Furthermore, no other plausible species are found 
at this wavelength in the 
NIST ASD or the \citet{kurucz11} line lists.
Unidentified lines at 2203.427 and 2203.645~\AA\
could be explained by Co~\textsc{ii} and V~\textsc{i}
transitions, respectively, only if the 
\loggf\ values recommended by the NIST ASD or the \citet{kurucz11} linelists
are underestimated by several dex.
We treat their strengths as free parameters in our analysis,
and this choice does not influence the derived Pb abundances.

The Pb~\textsc{ii} line is on the linear part of the curve-of-growth 
in \hdnine\ and \hdtwo, but it is saturated in \hdone.
We derive abundances 
using the 2017 version of the 
LTE line analysis software MOOG
\citep{sneden73,sobeck11}.
We adopt an \spro\ mix of Pb isotopes \citep{sneden08}
for \hdnine\ and \hdone\
and an \rpro\ mix for \hdtwo.
We generate the line list based on 
\citet{kurucz11},
\citet{peterson17}, and
the NIST ASD.~
We match the synthetic spectra to the observed spectra
following the general methods described by \citet{roederer12d}.

Table~\ref{abundtab} lists the derived abundances.
The Pb abundance is defined as 
\logeps{Pb}~$\equiv \log_{10}(N_{\rm Pb}/N_{\rm H})+$12.0.
The abundance ratio of elements Pb and Fe relative to the
Solar ratio is defined as
[Pb/Fe] $\equiv \log_{10} (N_{\rm Pb}/N_{\rm Fe}) - \log_{10} (N_{\rm Pb}/N_{\rm Fe})_{\odot}$, where
\logeps{Pb}$_{\odot} =$ 2.04 and \logeps{Fe}$_{\odot} =$ 7.50.
Following
\citet{roederer18c},
we compute 1$\sigma$ uncertainties by drawing $10^{3}$ resamples of
the stellar parameters,
\loggf\ values, and
equivalent widths approximated from 
the abundance derived via synthesis using a reverse
curve-of-growth method.

\subsection{The Pb Isotope Mix}
\label{isotopes}

\begin{figure}
\begin{center}
\includegraphics[angle=0,width=3.35in]{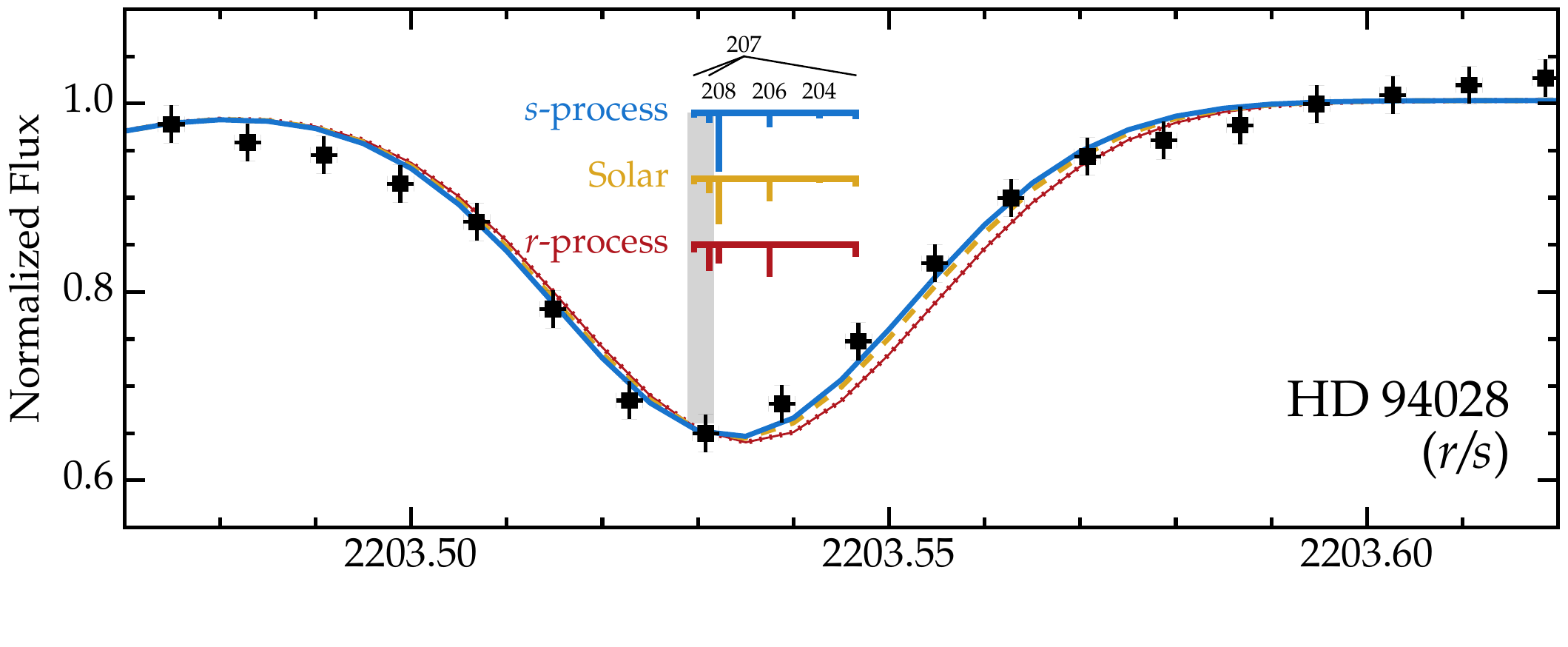} \\
\vspace*{-0.1in}
\includegraphics[angle=0,width=3.35in]{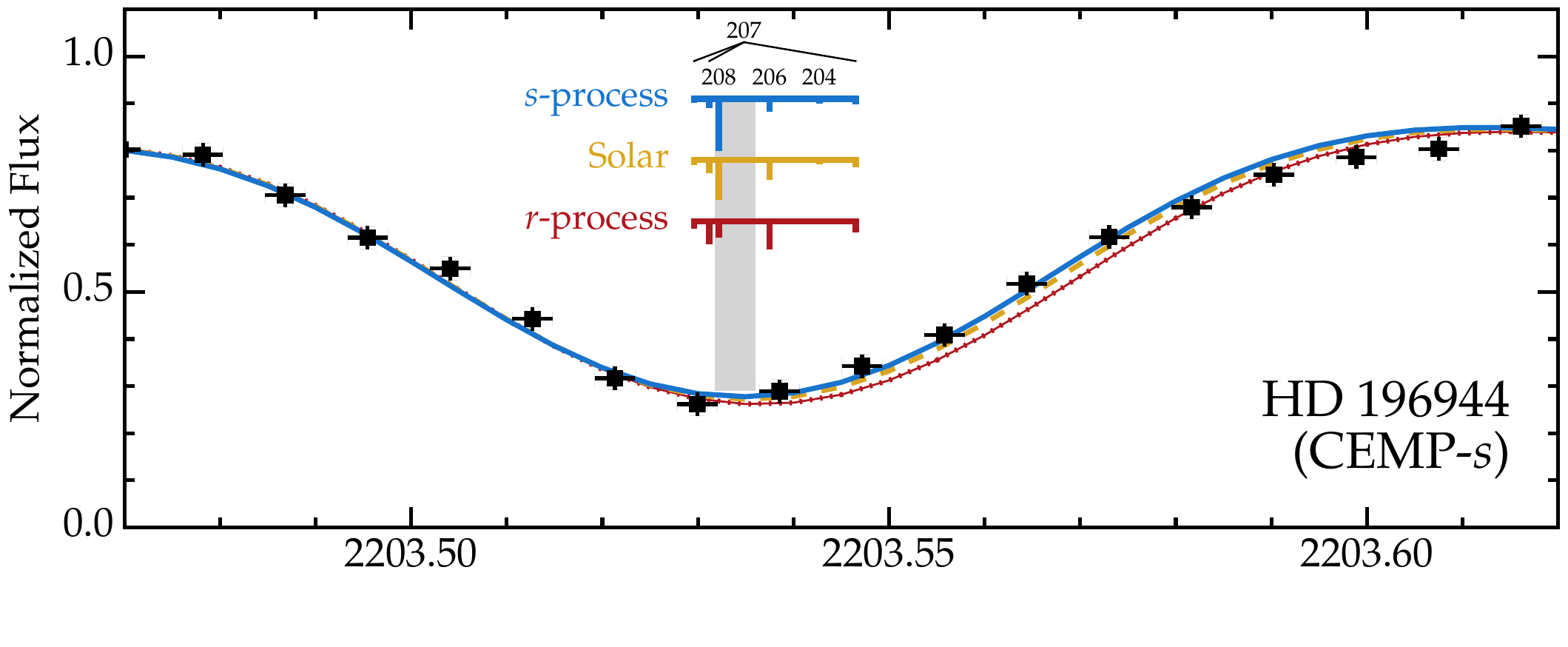} \\
\vspace*{-0.1in}
\includegraphics[angle=0,width=3.35in]{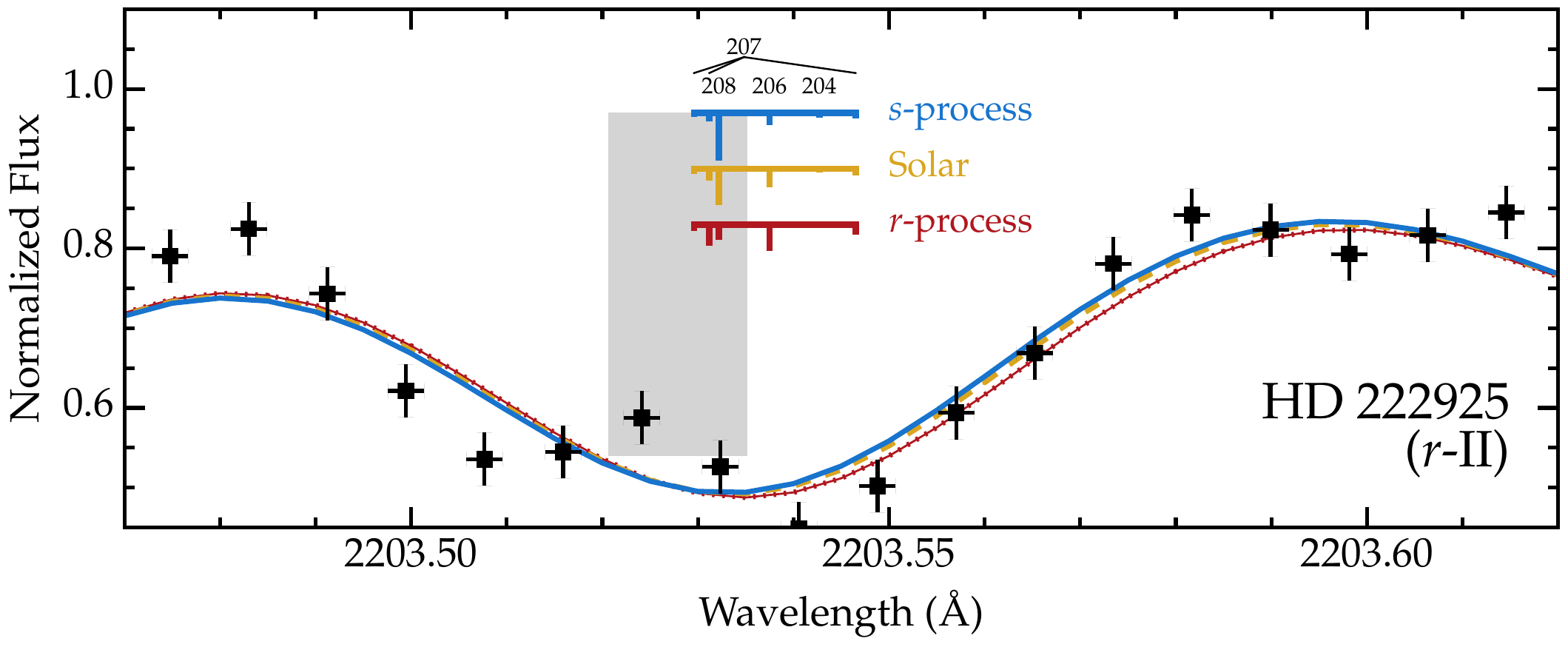}
\end{center}
\caption{
\label{plotiso}
Tight zoom around the Pb~\textsc{ii} line.
The Pb abundance is held fixed and the 
isotope mixes are varied.
The blue solid line represents an \spro\ isotope mix
($f_{204}/f_{206}/f_{207}/f_{208}$ =
0.025/0.143/0.137/0.695) from \citet{sneden08},
the yellow dashed line represents the Solar isotope mix
(0.014/0.241/0.221/0.524) \citep{meija16}, and
the red studded line represents an \rpro\ isotope mix
(0.000/0.359/0.451/0.190) from \citeauthor{sneden08} 
The filled dots represent the observed spectrum.
The shaded gray box in each panel represents the 
$\pm 1\sigma$ line centroid of the
Pb~\textsc{ii} line in the observed spectrum.
The isotope mixes inferred by our analysis (Section~\ref{isotopes}) 
are insensitive to the initial isotope mixes assumed for each star 
(Section~\ref{pbabunds}) once the abundances have been fixed.
}
\end{figure}

The Pb isotope mix has not been assessed previously 
in any metal-poor star.
The HFS of the $^{207}$Pb isotope,
particularly the upper level of the line at 2203~\AA,
and the IS
of the four Pb isotopes are wide compared to the
width of the stellar line profiles shown in Figure~\ref{plotiso}.
As the isotope mix shifts from the
\rpro, where the $^{206}$Pb isotope (35.9\%) 
and wide HFS of the $^{207}$Pb isotope (45.1\%) are expected to dominate,
to an \spro\ mix, where the $^{208}$Pb isotope 
is expected to dominate (69.5\%),
the absorption line profile narrows and shifts 
to shorter wavelengths.

The observed Pb~\textsc{ii} line profiles in
\hdnine\ and \hdone\ both favor a narrower profile,
suggesting that an even-$A$ isotope dominates.
We evaluate which isotope it might be
by comparing the line centroid to the isotope wavelengths.
We set the local wavelength zeropoint of the observed spectrum
relative to the synthetic spectra
using three relatively unblended Fe~\textsc{ii} lines,
with wavelengths known to better than
0.0011~\AA\ \citep{nave13},
located in the same echelle order
that contributes most of the signal to the Pb~\textsc{ii} line.
We also account for uncertainty in the
center-of-gravity wavelength of the
Pb~\textsc{ii} line (0.0007~\AA; \citealt{wood74})
and measurement uncertainties in the 
Fe~\textsc{ii} and Pb~\textsc{ii} line centroids.
As shown in Figure~\ref{plotiso},
the centroid of the observed Pb~\textsc{ii} line in 
\hdnine\ and \hdone\
favors absorption by the isotopes situated farthest to the blue, 
$^{207}$Pb and $^{208}$Pb.
Our method of setting the wavelength zeropoint
using Fe~\textsc{ii} lines appears to have 
produced a small mismatch between the observed and synthetic 
line profiles in \hdnine\ and \hdtwo.
Manual adjustment ($\leq 1\sigma$) 
of the observed spectrum in each case
so that it matches the blue side of the line,
which is relatively insensitive to the Pb isotope mix,
still favors the $^{207}$Pb and $^{208}$Pb isotopes.

We conclude from these two tests---the narrow line profiles 
and the positions of the line centroids---that
the $^{208}$Pb isotope is dominant
in \hdnine\ and \hdone.
The S/N is too low to draw any conclusions about \hdtwo.

\section{Discussion}
\label{discussion}

\subsection{Neutral Pb and Non-LTE Effects}
\label{nlte}

The abundances derived from the Pb~\textsc{ii} line are
higher than the abundances or upper limits derived from the Pb~\textsc{i} lines
in all three stars:\
[Pb~\textsc{ii}/Pb~\textsc{i}] = $+$0.36 $\pm$~0.34, 
$+$0.49 $\pm$~0.28, and $>+$0.04 
in \hdnine, \hdone, and \hdtwo, respectively.
\citet{mashonkina12} computed non-LTE corrections to the LTE
abundances for several metal-poor atmospheres.
That study found that the lower levels of 
the Pb~\textsc{i} lines at 2833 and 4057~\AA\
experience similar deviations from LTE.~
The ground state of singly-ionized Pb
is well-described by LTE in their calculations.
We assume that the excited 6$s^{2}$6$p$ level
that gives rise to the Pb~\textsc{ii} line at 2203~\AA\ is
also well-described by LTE.~
The \citeauthor{mashonkina12}\ non-LTE corrections for red giants
(dwarfs)
range from $+$0.26 to $+$0.62~dex ($+$0.22 to $+$0.32~dex)
when using a Drawin scaling factor, $S_{H}$, of 0.1,
which relates to the strength of inelastic collisions with 
neutral hydrogen.
No models in their grid exactly match 
the stars in our sample,
but the closest models predict non-LTE corrections 
$\approx +$0.27 and $+$0.52~dex for \hdnine\ and \hdone, respectively,
which match the offsets we derive in LTE.~
We support the conclusion of \citeauthor{mashonkina12}\
that departures from LTE impact Pb abundances derived
from Pb~\textsc{i} lines.

Pb is often used to constrain the \spro\ or \ipro\ models 
used to explain nucleosynthesis patterns in stars 
(e.g., \citealt{hampel19}).
A change in [Pb/Fe] by $+$0.4~dex is significant 
and could affect the final neutron exposure inferred from models,
which sets, for example, the estimated timescale for an \ipro\ event.
Future work should incorporate
non-LTE corrections to abundances derived from Pb~\textsc{i} lines
or derive Pb abundances in LTE directly from the UV Pb~\textsc{ii} line.

\subsection{Pb in the \spro}
\label{spropb}

Our LTE results confirm the enhanced Pb abundances in 
the \spro-enhanced stars \hdnine\ and \hdone.
We derive [Pb/Ba] = $+$1.05 $\pm$~0.35 in \hdone.
This is in good agreement with the AGB \spro\ model
of \citet{bisterzo10}
discussed at length in \citet{placco15cemps}
(see also \citealt{abate15}).
We derive [Pb/Ba] = $+$0.42 $\pm$~0.17 in \hdnine,
which supports the interpretation
of \citet{roederer16c} that 
the Ba and Pb in \hdnine\ originated mainly via the \spro.
Our isotopic analysis 
reaffirms theoretical predictions
that the large Pb overabundances in low-metallicity \spro\ environments
are dominated by $^{208}$Pb.

\subsection{Pb in the \rpro}
\label{rpropb}

The \logeps{Pb/Eu} ratio in \hdtwo, 0.76 $\pm$~0.14,
matches
the Solar System \rpro\ ratio,
\logeps{Pb/Eu} = 0.76 $\pm$~0.10
\citep{sneden08,bisterzo14,prantzos20}.
This result indicates that the \rpro\ residuals for the
Pb isotopes are, in aggregate, correct
when the effects of low-metallicity AGB stars
(i.e., the so-called ``strong component'') are included.
\citet{clayton67} argued that 
the dominant \rpro\ isotopes of Pb must be $^{206}$Pb and $^{207}$Pb.
The Pb~\textsc{ii} line centroid in \hdtwo\ is 
not in conflict with
this reasoning, although the S/N is too low in our spectrum
to support a more definitive statement.

The close coupling between Th and Pb enables
the use of Th/Pb as a
chronometer pair that is relatively
insensitive to the details of the \rpro\ model used to 
calculate the initial production ratio.
\hdtwo\ does not exhibit a prominent actinide boost \citep{roederer18c},
and its \logeps{Th/Pb} ratio is $-$1.20 $\pm$~0.14.
The \logeps{Th/Pb} ratio in \hdtwo\
corresponds to an age of 
8.2 $\pm$~5.8~Gyr
using the production ratios from \citet{roederer09b}.
The permitted age range is large, but
improving the S/N at the Pb~\textsc{ii} line
in future observations
would improve the age precision.
Pb lines are easier to detect in metal-poor stars
than the U~\textsc{ii} line at 3859~\AA,
and the Th/Pb chronometer offers 
an alternative model-insensitive age indicator to
the U/Th chronometer in \rpro-enhanced stars.

\acknowledgments

We thank E.A.\ Den Hartog for useful discussions
and the referee for a quick and helpful report.
I.U.R., J.E.L., T.C.B., A.F., and V.M.P.\
acknowledge support provided by NASA through grants
GO-14765 and GO-15657
from STScI,
which is operated by the AURA
under NASA contract NAS5-26555.
I.U.R., T.C.B., R.E., A.F., E.M.H., and V.M.P.\
acknowledge financial support from
grant PHY~14-30152 (Physics Frontier Center/JINA-CEE)
awarded by the U.S.\ National Science Foundation (NSF).~
We acknowledge additional support from NSF grants
AST-1716251 (A.F.) and
AST-1815403 (I.U.R.).
T.T.H.\ acknowledges generous support from the 
George P.\ and Cynthia Woods Institute for Fundamental Physics and Astronomy
at Texas A\&M University.
Parts of this research were supported by the 
Australian Research Council Discovery Project scheme (DP170100521) 
and Centre of Excellence for All Sky Astrophysics in 3 Dimensions (ASTRO 3D), 
through project number CE170100013.
This research has made use of NASA's
Astrophysics Data System Bibliographic Services;
the arXiv pre-print server operated by Cornell University;
the SIMBAD and VizieR
databases hosted by the
Strasbourg Astronomical Data Center;
the ASD hosted by NIST;
the MAST at STScI; 
and the
Image Reduction and Analysis Facility (IRAF) software packages.

\facility{HST (STIS), Smith (Tull Coud\'{e})}

\software{IRAF \citep{tody93},
matplotlib \citep{hunter07},
MOOG \citep{sneden73},
numpy \citep{vanderwalt11},
R \citep{rsoftware}}

\bibliographystyle{../aasjournal}

%
%

\end{document}